\documentclass[epj]{svjour}
\usepackage{graphics}
\usepackage{graphicx}
\usepackage{amssymb,amsmath,pifont}
\usepackage{breqn}
\usepackage{widetext}
\begin{document}
\title{Bulk viscosity, interaction and the viability of phantom solutions}
\author{Yoelsy Leyva \inst{1} \and Mirko Sep\'ulveda
}                     
\offprints{}          
\institute{Departamento de F\'isica, Facultad de Ciencias, Universidad de Tarapac\'a, Casilla 7-D, Arica, Chile 
}
\date{Received: date / Revised version: date}
%
\abstract{
We study the dynamics of a bulk viscosity model in the Eckart approach for a spatially flat Friedmann-Robertson-Walker (FRW) universe. We have included radiation and dark energy, assumed as perfect fluids, and dark matter treated as an imperfect fluid having bulk viscosity.  We also introduce an interaction term between the dark matter and dark energy components. Considering that the bulk viscosity is proportional to the dark matter energy density and imposing a complete cosmological dynamics, we find bounds on the bulk viscosity in order to reproduce a matter-dominated era (MDE). This constraint is independent of the interaction term. Some late time phantom solutions are mathematically possible. However, the constraint imposed by a MDE restricts the interaction parameter, in the phantom solutions, to a region consistent with a null value, eliminating the possibility of late time stable solutions with $w<-1$. From the different cases that we study, the only possible scenario, with bulk viscosity and interaction term, belongs to the quintessence region.  In this latter case, we find bounds on the interaction parameter compatible with latest observational data.
\PACS{
      {98.80.-k}{Cosmology}   \and 
      {98.80.Jk}{Mathematical and relativistic aspects of cosmology} \and
      {95.35.+d}{Dark matter}   \and 
      {95.36.+x}{Dark energy}
      } 
} 
\maketitle
\section{Introduction}
\label{intro}
Since the discovered of the present stage of acceleration of the Universe \cite{Riess1998a,Perlmutter1999b} many 
candidates have been proposed to explain such observational result \cite{Copeland2006,Buchert2008,Clifton2012,Mortonson2013,Koyama2016}.  Among them, the cosmological constant, $w_{\Lambda}=-1$, remains not only as the simplest alternative but also as consistent with the latest observational data \cite{Adeothers2016}. Despite this, the $\Lambda$CDM model is not able to explain those results that still point to a phantom Universe \cite{Adeothers2016}, $w<-1$. 

An interesting way to recover accelerated solutions is by introducing dissipative processes in ordinary fluids. This approach has been explored in literature through the modeling of bulk viscosity in ordinary matter fluids \cite{Barrow1986d,Waga1986,Maartens1995b,Bastero-Gil2012,Setare2014d,Li2009,Avelino2009,Velten2011,VeltenSchwarz2012,VeltenSchwarzFabrisEtAl2013,Velten2013,AvelinoLeyvaUrena-Lopez2013,CruzLepeLeyvaEtAl2014,Acquaviva2014,Acquaviva2015,Acquaviva2016a} in the context of Eckart \cite{Eckart1940a} or  lineal\cite{Israel1979} and non lineal\cite{Maartens1997,Chimento1997} Israel-Stewart theories.  

Following the dissipative approach, in \cite{Velten2013} was shown that phantom solutions can be obtained by accepting the existence of bulk viscosity whithin the Eckart theory in the $\Lambda$CDM model\footnote{Either the bulk viscosity was acting on the radiation, or on the pressureless matter, the crossing of the phantom divide is possible.}. This result was obtained by using multiple observational tests and considering that the bulk viscosity of some fluid depends on its own energy density, namely $\zeta_j=\zeta_j(\rho_j)$. This ansatz avoid the degeneracy problem associate when the bulk viscosity is taken as  $\zeta_j=\zeta_j(H)$\cite{Velten2013}.  However, in \cite{CruzLepeLeyvaEtAl2014}, the same scenario was studied, from the dynamical system point of view, finding that viscous phantom solutions with stable behavior are not allowed in the framework of complete cosmological dynamics \cite{AvelinoLeyvaUrena-Lopez2013,LeonLeyvaSocorro2014}. In the present paper we work along these lines by including an interaction between the dark matter and the dark energy. This kind of interaction mechanism has shown to be compatible with the current data \cite{Wang2016}. In the context of viscous fluids, the interaction between dark matter and dark energy was studied in \cite{AvelinoLeyvaUrena-Lopez2013}. It has been shown that, under the ansatz $\zeta_j=\zeta_j(H)$, low-redshift data favors a positive definite value of the bulk viscosity whereas, that high-redshift data prefers negative value of the bulk viscosity. This latter result is in tension with the local second law of thermodynamics (LSLT) \cite{Maartens1996,Zimdahl2000}, which it states that for an expanding universe $\zeta\geq0$\cite{Misner1973}.  

In the present work we are interested in extend the results obtained in \cite{Velten2013,CruzLepeLeyvaEtAl2014} by taking into account an interaction term between dark energy and dark matter and, at the same time, extend the results in \cite{AvelinoLeyvaUrena-Lopez2013} by exploring a different functional form for the bulk viscosity\footnote{Recall that in \cite{AvelinoLeyvaUrena-Lopez2013} the ansatz $\zeta_j=\zeta_j(H)$ was used, whereas in \cite{Velten2013,CruzLepeLeyvaEtAl2014} the bulk viscosity was taken as $\zeta_j=\zeta_j(\rho_j)$ in orden to avoid the model degeneration.}. 

The paper is organized as follow: in Section \ref{model} we present the field equation of the model. We take into account the contribution of pressureless matter,  radiation and dark energy. The first matter fluid is considered as an imperfect fluid, having bulk viscosity in the framework of the Eckart theory \cite{Eckart1940a}, whereas the remaining fluid obeying barotropic equation of state (EOS). The bulk viscosity coefficient is taken to be proportional to the dark matter energy density\footnote{More specifically  $\zeta\propto\rho_m^{\frac{1}{2}}$.}. In Section \ref{tas}, we study the evolution of the field equations from the perspective of the equivalent autonomous system. We focus our attention in a particular form for the interaction term between the dark matter and dark energy components. A detailed discussion about the viability  of a complete cosmological dynamics \cite{AvelinoLeyvaUrena-Lopez2013,LeonLeyvaSocorro2014} is provided. Important constraints on the bulk viscosity and interaction parameter are obtained. Finally, Section \ref{conclusion} is devoted to conclusions.

\section{The model}\label{model}
We study a cosmological model in a spatially flat FRW background metric, in which the matter components are radiation, dark matter and dark energy. We assumed that the dark matter fluid presents bulk viscosity in the framework of  the Eckart theory, whereas the radiation and dark energy are assumed as perfect fluids. Following this set up, the Friedmann constraint, the conservation equations for the matter fluids and the Raychadury equation can be written as:
\begin{eqnarray}
    3H^2 &=& \left( \rho_{\rm r}  + \rho_{\rm dm} +\rho_{\rm de} \right), \label{ConstrainFriedmann} \\
    \dot{\rho}_{\rm r}  &=& - 4 H \rho_{\rm r}, \label{EqConsRad} \\
    \dot{\rho}_{\rm dm} &=& - 3 H\rho_{\rm dm}+9H^2 \zeta+Q, \label{EqConsDM} \\
   \dot{\rho}_{\rm de} &=& - 3H\gamma_{\rm de}\rho_{\rm de}-Q,  \label{EqConsDE}\\
    \dot{H} &=& -\frac{1}{2} \left( \rho_{\rm dm}  +\frac{4}{3} \rho_{\rm r} +\gamma_{\rm de} \rho_{\rm de} -3H\zeta \right), \label{eq:Ra}
  \end{eqnarray}
where $G$ is the Newton gravitational constant, $H$ the Hubble parameter,
$(\rho_{\rm dm}$, $\rho_{\rm r}$, $\rho_{\rm de})$ are the energy
densities of dark matter, radiation and  DE fluid components respectively. Whereas, $\gamma_{\rm de}$ is the barotropic index of the EOS
of DE, which is defined from the relationship $p_{\rm de} = (\gamma_{\rm de} -1)
\rho_{\rm de}$, where  $p_{\rm de}$ is the pressure of DE. 
The term $Q$ in (\ref{EqConsDM}-\ref{EqConsDE}) is the interaction term between the dark matter and the dark energy components, while $9H^2\zeta$ in Eq.~(\ref{EqConsDM}) corresponds to the
bulk viscous pressure of the dark matter fluid, with $\zeta$ the bulk viscous coefficient.

We assume the bulk viscous coefficient $\zeta$ to be proportional to the energy density of the dark matter component in the form:
\begin{equation}
  \label{ViscosityDefinition}
 \zeta= \xi  \left(\frac{\rho _{dm}}{\rho _{dm0}}\right)^{\frac{1}{2} },
\end{equation}
where $\rho_{dm 0}$ is the present day value of the dark matter energy density.

\section{The autonomous system}\label{tas}

In order to study the dynamical properties of the system (\ref{EqConsRad}-\ref{EqConsDE},\ref{eq:Ra}), we introduce the following dimensionless phase space variables to build an autonomous dynamical system:
\begin{equation}\label{eq:R}
 x=\Omega_{de}\equiv\frac{\rho_{\rm de}}{3H^{2}} , \quad y = \Omega_{dm}\equiv\frac{ \rho_{\rm dm}}{3H^{2}}, \quad \Omega_r\equiv\frac{\rho_{\rm r}}{3H^{2}},
\end{equation}
using the Friedmann constraint (\ref{ConstrainFriedmann}) is possible to reduce one degree of freedom, namely $\Omega_r=1-x-y$. Then the equation of motion can be written as
\begin{dmath}\label{a1}
\frac{dx}{dN}=3 x^2 \gamma _{\text{de}}-3 x \gamma _{\text{de}}-4 x^2-3 \xi _0 x \sqrt{y}-x y+4 x-\frac{Q}{3 H(t)^3},
\end{dmath}
\begin{dmath}\label{a2}
\frac{dx}{dN}=3 x y \gamma _{\text{de}}-4 x y-3 \xi _0 y^{3/2}-y^2+3 \xi _0 \sqrt{y}+y+\frac{Q}{3 H(t)^3},
\end{dmath}
where the derivatives are with respect to the $e$-folding number $N\equiv \ln a$  and we have introduced the dimensionaless parameter 
\begin{equation}
\xi=\frac{\xi }{H_0 \sqrt{\Omega _{10}}},
\end{equation}
where, in order to guarantees nonviolation of the LSLT\cite{Misner1973,Maartens1996,Zimdahl2000}, $\xi>0$\footnote{We are not taking into account the value $\xi=0$, that obviously fulfill the LSLT, because we are interested in studying  the effects of bulk viscosity in this set up. }

In addition, in orden to achieve an autonomous system from (\ref{a1}-\ref{a2}) we must define the interaction function $Q$. If the interaction term is taken as $Q=3Hf(\rho_m,\rho_{de})$\cite{Quartin2008,Caldera-Cabral2009a,Zimdahl2001,Zimdahl2003,Guo2007,AvelinoLeyvaUrena-Lopez2013,Wang2016}, then we can introducce a new function
\begin{equation}
\quad z \equiv\frac{Q}{3 H^3}=z(x,y),
\end{equation}
hence, the system (\ref{a1}-\ref{a2}) can be written as a two-dimensional autonomous system
\begin{equation}
\frac{dx}{dN}=3 (x-1) x \gamma _{\text{de}}-3 \xi _0 x \sqrt{y}-x(4 x+y-4)-z,\label{as1}
\end{equation}
\begin{equation}
\frac{dy}{dN} = 3 x y \gamma _{\text{de}}-y (4 x+y-1)-3 \xi _0 (y-1) \sqrt{y}+z. \label{as2}
\end{equation}
Imposing the conditions that radiation, dark matter and DE components be positive, definite, and  bounded at all times, we can define the phase space of Eqs. (\ref{as1}-\ref{as2}) as
\begin{equation}
  \Psi =\{(x,y): 0 \leq 1-x-y \leq 1 ,  0\leq x \leq1 , 0 \leq y \leq 1\} \, . \label{eq:space}
\end{equation}
Moreover, we can introduce other cosmological parameter of interest, like the deceleration parameter ($q=-(1+\dot{H}/H^2)$) and the total effective EOS ($w_{eff}$) in terms of the dimensionless phase space variables (\ref{eq:R}):
\begin{eqnarray}
  q &=&\frac{1}{2} \left(3 x \gamma _{\text{de}}-4 x-3 \xi _0 \sqrt{y}-y+2\right)\, \label{q},\\
  w_{eff} &=& \frac{1}{3} \left(3 x \gamma _{\text{de}}-4 x-3 \xi _0 \sqrt{y}-y+1\right)\label{w}.
\end{eqnarray}

\begin{center}
\begin{table*}[t!]\caption[Qf]{Location, existence conditions
    according to the phase space (\ref{eq:space}), and
    stability of the critical points of the autonomous system
    (\ref{as1})-(\ref{as2}) for $\gamma_{de}=0$ and $z=3\beta x$. The
    eigenvalues of the linear perturbation matrix associated to each
    of the following critical points are displayed in
    Table~\ref{tab2}. We have introduced the definitions $A=\sqrt{4 \beta +\xi _0^2}$, $B=\sqrt{-A \xi _0+2 \beta +\xi _0^2}$, $C=\sqrt{A \xi _0+2 \beta +\xi _0^2}$ and $D=8-8\beta-\zeta_{0}^2$.}
  \begin{tabular}{@{\extracolsep{\fill}}| c | c | c | c | c | c |}
    \hline\hline
    $P_i$ & $x$ & $y$ & Existence & Stability \\
    \hline\hline
   $P_{1}$ & $0$ & $0$ & Always & Unstable if $\beta<\frac{4}{3}$ \\
       & & & & Saddle  if $\beta>\frac{4}{3}$  \\ \hline
	        	          
$P_{2}$ & $0$ & $1$ & Always & Saddle if $\beta <1\land 0<\xi _0<1-\beta$ \\
& & & & Stable if $\left(\beta \leq 1\land \xi _0>1-\beta \right)\lor $       \\   
& & & & $\left(\beta >1\land \xi _0>0\right)$       \\     
	\hline
$P_{3}$ & $\frac{1}{8} \left(\xi _0 \left(A-3 \sqrt{2} B\right)+D\right)$ & $\frac{1}{2} \left(-A \xi _0+2 \beta +\xi _0^2\right)$ & $\beta \leq 0\land$ & See discusion in subsection \ref{subscp} \\
	      		&     &     &  $\left(4\beta +\xi _0^2\geq 0\land 0<\xi _0\leq 2\right)\lor$      & \\
			 &     &     &  $\left(\xi _0>2\land \beta +\xi _0\geq 1\right)$      & \\\hline
			
$P_{4}$ & $\frac{1}{8} \left(-\xi _0 \left(A+3 \sqrt{2} C\right)+D\right)$ & $\frac{1}{2} \left(A \xi _0+2 \beta +\xi _0^2\right)$ & $\left(0<\xi _0<2\land -\frac{\xi _0^2}{4}\leq \beta \leq 1-\xi _0\right)\lor$ & See discusion in subsection \ref{subscp}\\
	      		&     &     &  $\left(\xi _0=2\land \beta =-1\right)$      &  \\\hline
   
\end{tabular}\label{tab1}
\end{table*}
\end{center}

\begin{center}
\begin{table*}[t!]\caption[Qf]{Eigenvalues and some basic physical
    parameters for the critical points listed in Table~\ref{tab1}, see
    also Eqs.~(\ref{q}) and~(\ref{w}). }
  \begin{tabular}{@{\extracolsep{\fill}}| c | c | c | c | c | c | c |}
    \hline \hline
    $P_i$&$\lambda_1$ & $\lambda_2$ & $\Omega_r$&$w_{eff}$ & $q$\\
    \hline\hline
  $P_{1}$ & $4-3 \beta$ & sgn($\xi$)$\infty$ & $1$ & $\frac{1}{3}$ & $1$ \\ \hline

  $P_{2}$ & $-3 \xi _0-1$ & $-3 \left(\beta +\xi _0-1\right)$ &$0$ &$-\xi _0$ &$\frac{1}{2} \left(1-3 \xi _0\right)$ \\ \hline
  
  $P_{3}$ &See appendix \ref{apen1} &See appendix \ref{apen1} & $0$& $-1+\beta$& $\frac{3 \beta }{2}-1$ \\ \hline
  
  $P_{4}$ &See appendix \ref{apen2} & See appendix \ref{apen2}&$0$ &$-1+\beta$ & $\frac{3 \beta }{2}-1$\\ \hline
  
   \hline
\end{tabular}\label{tab2}
\end{table*}
\end{center}

\subsection{Dynamics of the autonomous system}
Despite the autonomous system (\ref{as1}-\ref{as2}) allows to study the dynamics of (\ref{EqConsRad}-\ref{EqConsDE},\ref{eq:Ra}) for general interaction functions of the form $z=z(x,y)$, we will focus our attention in the particular case of $z=3\beta H \rho_{de}=3 \beta x$\cite{Quartin2008,Caldera-Cabral2009a,Kremer2012,Wang2016}. The choice of this particular form is motivated by the requirement of a critical point associated with a MDE in order to explain the structure formation. A simple inspection of (\ref{as1}-\ref{as2}) shows that  the latter requirement implies that $(x,y)=(0,1)$ thus $z(0,1)=0$. Hence, only those interaction function that fulfill this condition are able to allow a MDE\footnote{See a similar analysis in the case of the ansatz $\zeta_i=\zeta_i(H)$ in \cite{AvelinoLeyvaUrena-Lopez2013}}. We also will restrict our analysis to the case $\gamma_{de}=0$. The full set of critical points of (\ref{as1}-\ref{as2}) are summarized in Table \ref{tab1}, whereas the corresponding eigenvalues of the linear perturbation matrix are given in Table \ref{tab2}

\subsubsection{Critical points and stability}\label{subscp}

$P_1$ represents a decelerating solution($q=1$, $w_{eff}=1/3$) dominated by the radiation component, $\Omega_r=1$, and exists unrestrictedly of the sing/value of the interaction and bulk viscosity parameters. However, its stability behavior depends on the value of the interaction parameter $\beta$, namely, 
(i) unstable if $\beta<4/3$ or (ii) saddle if $\beta>4/3$.

Critical point $P_2$ corresponds to a pure dark matter-domination period ($\Omega_m=1$) and always exists. \footnote{Recall that $P_3$, like $P_1$, exists independently of the values/sign of the bulk viscosity and the existence of interaction between the dark components. This point has a similar behavior that points $2a$ in \cite{AvelinoLeyvaUrena-Lopez2013} and $P_2$ in \cite{CruzLepeLeyvaEtAl2014}} If the condition $\xi_0^2\ll0$ is satisfied, then this point corresponds to the standar matter-domination period, namely $w_{eff}\approx0$ and $q\approx1/2$. Otherwise $w_{eff}$ is negative and can behave as an acelerated solution if $\xi _0>\frac{1}{3}$ or even as a phantom solution if $\xi _0>1$.  As Table \ref{tab1} and \ref{tab2} show, these acelerated solutions are possible in absence of dark energy ($x=\Omega_{de}=0$). From the stability point of view, $P_3$ displays two different behaviors, that is: i) saddle if $\beta <1\land 0<\xi _0<1-\beta$ or ii) stable if $\left(\beta \leq 1\land \xi _0>1-\beta \right)\lor $ $\left(\beta >1\land \xi _0>0\right)$.

\begin{figure}[tbp]
\resizebox{0.49\textwidth}{!}{\includegraphics{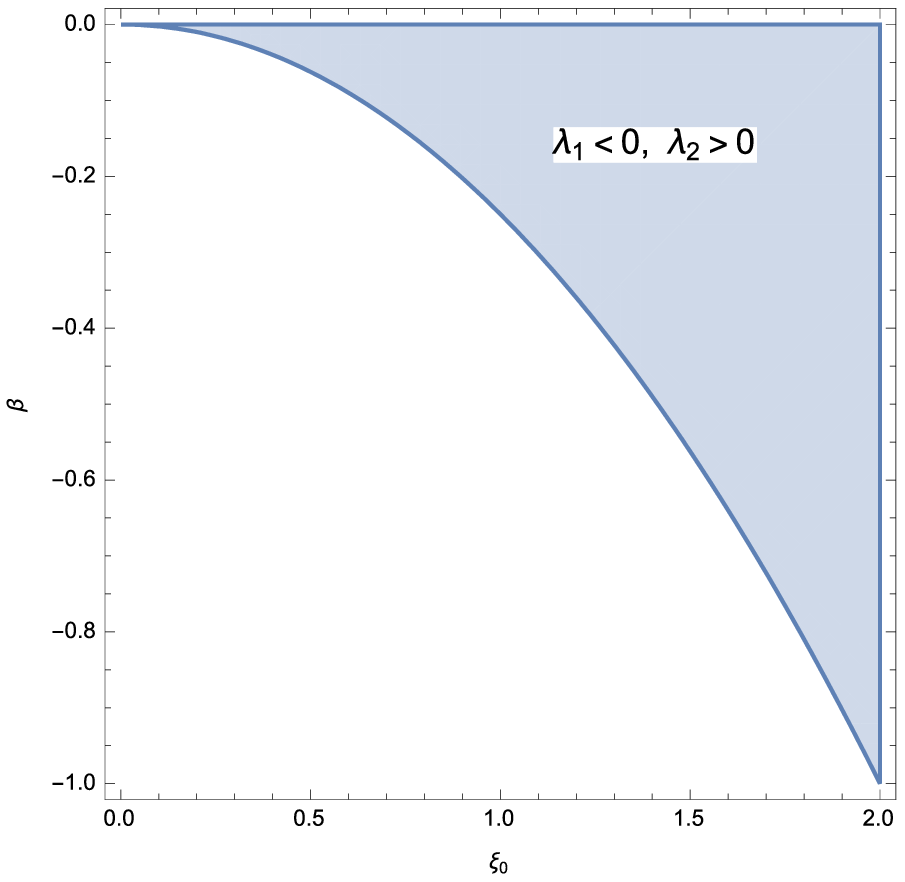}}
\caption{Saddle ($\lambda_1<0$ and $\lambda_2>0$) region for $P_3$ in the phantom region case $1$. See the corresponding eigenvalues($\lambda_1$, $\lambda_2$) in appendix \ref{apen1}}
\label{fig1} 
\end{figure}

\begin{figure}[tbp]
\resizebox{0.49\textwidth}{!}{\includegraphics{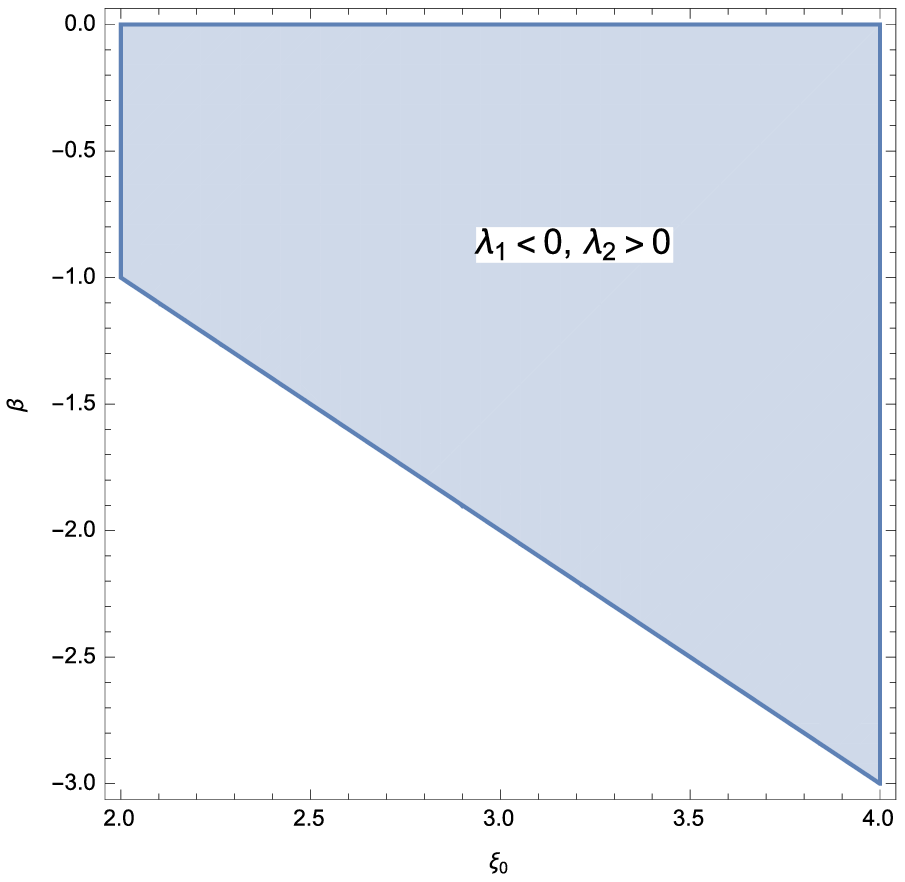}}
\caption{Saddle ($\lambda_1<0$ and $\lambda_2>0$) region for $P_3$ in the phantom region case $2$. See the corresponding eigenvalues($\lambda_1$, $\lambda_2$) in appendix \ref{apen1}}
\label{fig2} 
\end{figure}

$P_3$ represents an scaling solutions between dark matter and dark energy components and exists when
\begin{equation}\nonumber
\left(\beta \leq 0\land 4 \beta +\xi _0^2\geq 0\land 0<\xi _0\leq 2\right)\lor
\end{equation}
\begin{equation}
\left(\beta \leq 0\land \xi _0>2\land \beta +\xi _0\geq 1\right)\nonumber
\end{equation}
A background level, $P_3$ is able to mimic accelerated solutions\footnote{Unlike the previous critical points ($P_1$-$P_2$), is not possible to reproduce, in the region of existence, decelerated solutions such as pressureless matter ($w_{eff}=0$) or radiation ($w_{eff}=1/3$).}  in the phantom and de Sitter regions, namely: \\
i) phantom region ($w_{eff}<-1$)
\begin{enumerate}
\item  Saddle if $0<\xi _0\leq 2\land -\frac{\xi _0^2}{4}\leq \beta <0$, see Fig. \ref{fig1} for more details.
\item  Saddle if $\xi _0>2\land 1-\xi _0\leq \beta <0$, see Fig. \ref{fig2} for more details.
\end{enumerate}
ii) de Sitter region ($w_{eff}=-1$)
\begin{enumerate}
\item Only if $\beta=0$. We  are not considered this case here because that means null interaction between dark matter and dark energy\footnote{The case with $\beta=0$ was studied in \cite{CruzLepeLeyvaEtAl2014}}.
\end{enumerate}
 
\begin{figure}[tbp]
\resizebox{0.49\textwidth}{!}{\includegraphics{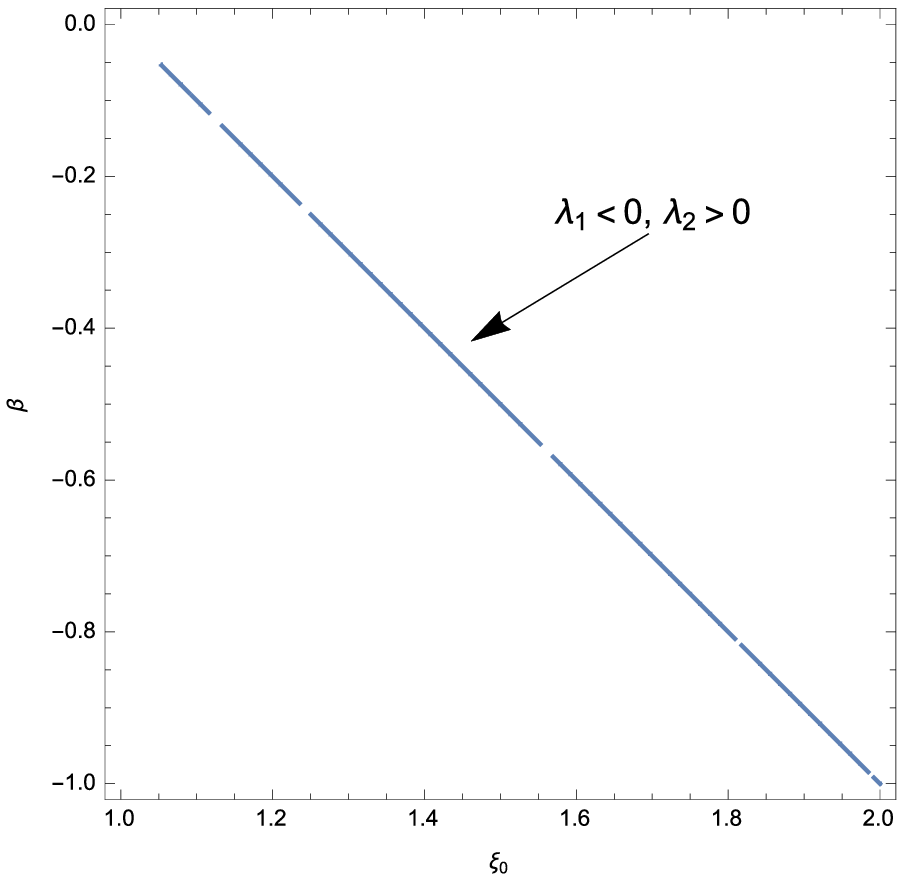}}
\caption{Saddle ($\lambda_1<0$ and $\lambda_2>0$) region for $P_4$ in the phantom region case $2$. See the corresponding eigenvalues($\lambda_1$, $\lambda_2$) in appendix \ref{apen1}}
\label{fig3}
\end{figure}
\begin{figure}[tbp]
\resizebox{0.49\textwidth}{!}{\includegraphics{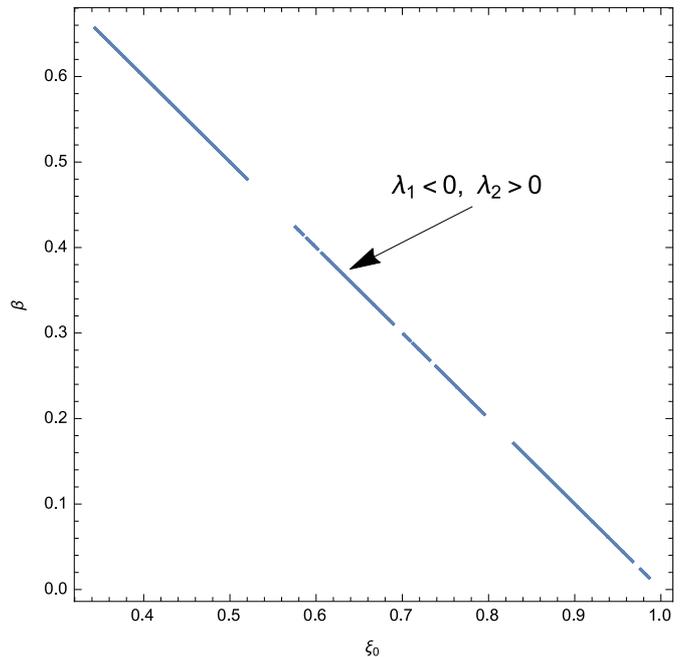}}
\caption{Saddle ($\lambda_1<0$ and $\lambda_2>0$) region for $P_4$ in the quintessence region case $2$. See the corresponding eigenvalues($\lambda_1$, $\lambda_2$) in appendix \ref{apen1}}
\label{fig4}
\end{figure}

Critical point $P_4$ corresponds to a scaling solution between dark matter and dark energy. This point exists in the region  
\begin{equation}\nonumber
\left(0<\xi _0<2\land -\frac{\xi _0^2}{4}\leq \beta \leq 1-\xi _0\right)\lor \left(\xi _0=2\land \beta =-1\right).
\end{equation}
In the existence regions, $P_4$ is able to mimic only accelerated solutions, namely\\
iii) phantom region ($w_{eff}<-1$)
\begin{enumerate}
\item Stable if $0<\xi _0\leq 1\land -\frac{\xi _0^2}{4}\leq \beta <0$
\item Saddle if $1<\xi _0<2\land -\frac{\xi _0^2}{4}\leq \beta \leq 1-\xi _0$ in a narrow region in the parameter space ($\xi_0$, $\beta$), as Fig. \ref{fig3} shows, otherwise is stable.
\item If $\xi_0=2\land\beta=-1$ then $w_{eff}=-2$, being an unrealistic value for the effective EOS parameter.
\end{enumerate}
iv) de Sitter region ($w_{eff}=-1$)
\begin{enumerate}
\item As in $P_3$, $\beta=0$ leads to a de Sitter solution. As we mention before, this is discarded because it requires a null interaction between dark matter and dark energy\footnote{Recall that the null interaction case was developed in \cite{CruzLepeLeyvaEtAl2014}}
\end{enumerate}
v) quintessence region ($-1<w_{eff}<-1/3$).
\begin{enumerate}
\item Stable if $0<\xi _0\leq \frac{1}{3}\land 0<\beta <\frac{2}{3}$
\item Saddle if $\frac{1}{3}<\xi _0<1\land 0<\beta \leq 1-\xi _0$ in a narrow region in the parameter space ($\xi_0$, $\beta$), as Fig. \ref{fig4} shows, otherwise is stable.
\end{enumerate}

\subsubsection{Cosmological evolution}
According to current observational data,  any model that aims to make a complete description of the evolution of the Universe must have to follow the complete cosmological paradigm (\cite{AvelinoLeyvaUrena-Lopez2013,CruzLepeLeyvaEtAl2014,LeonLeyvaSocorro2014}). This paradigm impose transitions between three different evolutions eras from early times to late times, namely: i) radiation-dominated era (RDE), ii) matter-dominated era (MDE) at intermediate stage of evolution, and iii) accelerated expansion. Every one of these statement can be translate into critical point connected by heteroclinic orbits\cite{2005,Coley2013,HeinzleUgglaRohr2009,Urena-Lopez2012}. 

\begin{figure}[tbp]
\resizebox{0.49\textwidth}{!}{
  \includegraphics{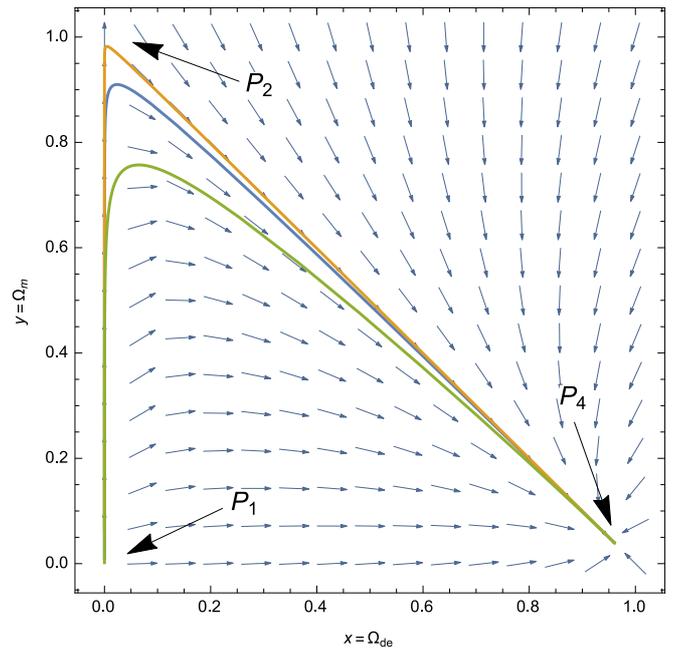}}
\caption{Vector field in the plane ($x$, $y$) for the autonomous system (\ref{as1})-(\ref{as2}) with $\gamma_{de}=0$. The free parameters have been chosen as ($\xi_0$, $\beta$)$=$($0.0008$, $0.038$). In this case, the quintessence solution, $P_{4}$, is the late time attractor of the system, representing an accelerated solution ($w_{eff}=-0.96$). The transition from the RDE ($P_{1}$) to $P_{4}$ allows for the selection of appropriate initial conditions to recover a true MDE ($P_{3}$) with $w_{eff}\simeq0$ according with condition (\ref{cond1})\cite{ThomasKoppSkordis2016}.}
\label{fig5} 
\end{figure}

The condition for a purely RDE ($\Omega_r=1$) is always satisfied by $P_1$, independently of the value of the bulk viscosity parameter $\xi_0$. Its unstable behavior, given that $\beta<4/3$, guarantees that it can be the source of any solution in the phase space. 

For intermediate stages of cosmic evolution, the presence of MDE is needed in order to  describe the formation of structures. This matter-dominated period can be recovered by $P_2$ . This critical point exists independently of the value of the bulk viscosity parameter  but a background level, for a no null value of $\xi_0$, it behaves as a decelerating solution\footnote{As the existence of this critical point is also independently of the interaction between dark matter and dark energy ($\beta=0$), this results recover the behavior of $P_2$ in \cite{CruzLepeLeyvaEtAl2014}} if $0<\xi_0<1/3$ but, if the bulk viscosity takes a sufficiently small value, $\xi_0\ll0$, is possible to recover $w_{eff}\approx0$ and $q\approx1/2$. However this latter statement may be in tension with the recents constraints to the dark matter EOS, which state that $-0.000896<w_{dm}<0.00238$ at the $3\sigma$ level \cite{ThomasKoppSkordis2016} using the lastest Planck data realease \cite{Adeothers2016}\footnote{Among others, similar constraints are found in \cite{Muller2005,BarrancoBernalNunez2015}}. Thus, only tiny contribution of bulk viscosity is allowed in order to recover a true MDE with $P_2$:
\begin{equation}\label{cond1}
0<\xi_0<0.000896,
\end{equation}
this constraints on $\xi_0$ are also consistent with those obtained in \cite{VeltenSchwarz2012,VeltenSchwarzFabrisEtAl2013} in the absence of interaction between dark matter and dark energy. As we mentioned in subsection \ref{subscp}, $P_2$ is able to reproduce an accelerated solution given that $\xi_0>1/3$. However, as Table \ref{tab1} and \ref{tab1} show, this possible behavior has to be rule out because of the impossibility of finding another critical point 

Concerning the late time evolution of the Universe, the model has two more critical point capable of providing accelerated solutions, namely $P_3$ and $P_4$. Both represent scaling solutions between dark matter and dark energy. As was discussed in the previous subsection \ref{subscp}, from the mathematical point of view, is possible to obtain phantom, de-Sitter and quintessence solutions with saddle or stable behaviors depending of the values of the free parameters ($\xi_0$, $\beta$). If the interaction parameter is negative ($\beta<0$), meaning an energy transfer from dark matter to dark energy, is possible to obtain a late time transition between two phantom solutions:  $P_3$ case $i)1$ (saddle)  $\rightarrow$ $P_4$ case $iii)1.$ (stable). This transition requires $-\xi_0^2/4\leq\beta<0$ but, if we also demand a previos stages of RDE and MDE we must impose condition (\ref{cond1}), leading to an almost null value for the interaction parameter
\begin{equation}\label{cond2}
-2.00704\ast10^{-7}<\beta <0,
\end{equation}
thus the phantom solutions $P_3$ and $P_4$ tends to de Sitter solutions $w_{eff}=-1$ ($\beta=0$). The rest of the late time phantom solutions demand very large values of the bulk viscosity parameter, $\xi_0>1$, compared to those allowed by (\ref{cond1}) in order to recover a true MDE, hence they are ruled out.

The only possible late time scenario with a non null value of the interaction parameter corresponds to a stable quintessence solution ($P_4$).  This solution requires 
\begin{equation}\nonumber
0<\xi _0\leq \frac{1}{3}\land 0<\beta <\frac{2}{3}
\end{equation}
If we impose the condition (\ref{cond1}) to ensure a true MDE and, take into account the latest constraint on the value of the dark energy EOS \cite{Adeothers2016}, the following tiny region is obtained for the interaction parameter
\begin{equation}\label{cond4}
0<\beta\leq0.039.
\end{equation}
Fig. \ref{fig5} shows some example orbits in the plane ($x=\Omega_{de}$, $y=\Omega_m$) to illustrate the above scenario.

\section{Concluding remarks}\label{conclusion}
In this work we studied the dynamics of model of the universe filled with radiation, dark matter and dark energy. The dark matter component was treated as an imperfect fluid having bulk viscosity whereas the  remaining fluids were considered as a perfect fluids. The bulk viscosity was taken as proportional to the dark matter density $\zeta\propto\rho_m^{\frac{1}{2}}$ \cite{Velten2013} and, we introduce an interaction term between the dark matter and the dark energy components with the objetive of extend previous results developed in \cite{CruzLepeLeyvaEtAl2014}. This new term was taken as $Q=3 H \rho_{de}$ \cite{Quartin2008,Caldera-Cabral2009a,Kremer2012,Wang2016}.

Recall that the ansatz on the bulk viscosity used in \cite{Kremer2012,AvelinoLeyvaUrena-Lopez2013} ($\zeta\propto H$) is different from the used in this work. Thus the results obtained now are new compared with those obtained in \cite{Kremer2012,AvelinoLeyvaUrena-Lopez2013} and extend those obtained in \cite{CruzLepeLeyvaEtAl2014} by the introduction of the interaction term. 

We performed a dynamical system analysis of the model in order to investigate its asymptotic evolution and behavior. The imposition of a transition from a RDE to an accelerated dominated solution, passing through a true MDE reduce the possible values of the bulk viscosity parameter to a tiny region $0<\xi_0<0.000896$. This finding is independent of the value of $\beta$ and support those obtained in \cite{VeltenSchwarz2012,VeltenSchwarzFabrisEtAl2013,CruzLepeLeyvaEtAl2014} with no interaction between dark matter and dark energy. The presence of an interaction between dark matter and dark energy allows, from the mathematical point of view, to obtain stable(saddle) late time accelerated solutions in the phantom, de Sitter and quintessence regions. However, the requirement of a true MDE imposes strong constraints on the interaction parameter $\beta$ in the case of late time  phantom solutions. In both cases, regardless of the direction of energy transfer between dark matter and dark energy, the interaction parameter is consistent with a null value, hence the de Sitter solution will be the late time attractor. Moreover, the impossibility of having late time accelerated solutions, caused solely by the viscous matter ($P_2$), found in \cite{CruzLepeLeyvaEtAl2014} with $\beta=0$, was extended to this new scenario with interaction between dark matter and dark energy.

The only favorable scenario with a no null value of the interaction parameter, $0<\beta\leq0.039$, is described by the late time stable quintessence solution $P_4$. This solution is able to fulfill the complete cosmological paradigm, that is a transition between $P_1$(RDE) $\rightarrow$ $P_2$ (MDE) $\rightarrow$$P_4$. Recall that this quintessence solution is compatible with the latest constraint on the values of the dark energy EOS \cite{Adeothers2016}.

\section{Acknowledgments}
This work was funded by Comisi\'on Nacional de
Investigaci\'on Cient\'{\i}fica y Tecnol\'ogica (CONICYT) through FONDECYT Grant $11140309$ (Y. L.).

\appendix

\section{Eigenvalues of critical point $P_4$}\label{apen1}


The eigenvalues of $P_4$ are:

\begin{widetext}
\begin{equation}
\lambda_1 = -\frac{\sqrt{-18 A \xi _0^5+3 F_5 \xi _0^4-6 F_4 \xi _0^3+6 F_3 \xi _0^2+24 F_2 \xi _0+8 F_1+9 \xi _0^6}}{4 \sqrt{2} B}-\frac{3 A \xi _0^2}{4 \sqrt{2}B}+3 \beta +\frac{3 \beta  \xi _0}{2 \sqrt{2} B}+\frac{3 \xi _0^3}{4 \sqrt{2} B}+\frac{3 \xi _0}{2 \sqrt{2} B}-\frac{7}{2}
\end{equation}
\begin{equation}
\lambda_2 = \frac{\sqrt{-18 A \xi _0^5+3 F_5 \xi _0^4-6 F_4 \xi _0^3+6 F_3 \xi _0^2+24 F_2 \xi _0+8 F_1+9 \xi _0^6}}{4 \sqrt{2} B}-\frac{3 A \xi _0^2}{4 \sqrt{2}B}+3 \beta +\frac{3 \beta  \xi _0}{2 \sqrt{2} B}+\frac{3 \xi _0^3}{4 \sqrt{2} B}+\frac{3 \xi _0}{2 \sqrt{2} B}-\frac{7}{2}
\end{equation}
\end{widetext}

where
\begin{equation}
A=\sqrt{4 \beta +\xi _0^2}
\end{equation}
\begin{equation}
B=\sqrt{-A \xi _0+2 \beta +\xi _0^2}
\end{equation}
\begin{equation}
F_1= (7-6 \beta )^2 B^2-24 \beta  \left(3 \beta ^2-7 \beta +4\right)
\end{equation}
\begin{equation}
F_2= A \left(15 \beta ^2-32 \beta +16\right)+\sqrt{2} \left(3 \beta ^2-3 \beta +1\right) B
\end{equation}
\begin{equation}
F_3=3 \sqrt{2} A (\beta -2) B-42 \beta ^2+124 \beta -58
\end{equation}
\begin{equation}
F_4=2 A (6 \beta -1)+3 \sqrt{2} (\beta -2) B
\end{equation}
\begin{equation}
F_5=3 A^2+24 \beta -4
\end{equation}

\section{Eigenvalues of critical point $P_5$}\label{apen2}

The eigenvalues of $P_5$ are:

\begin{widetext}
\begin{equation}
\lambda_1 =-\frac{\sqrt{18 A \xi _0^5+3 F_{10} \xi _0^4+6 F_9 \xi _0^3-6 F_8 \xi _0^2-24 F_7 \xi _0+8 F_6+9 \xi _0^6}}{4 \sqrt{2} C}+\frac{3 A \xi _0^2}{4 \sqrt{2}C}+3 \beta +\frac{3 \beta  \xi _0}{2 \sqrt{2} C}+\frac{3 \xi _0^3}{4 \sqrt{2} C}+\frac{3 \xi _0}{2 \sqrt{2} C}-\frac{7}{2}
\end{equation}
\begin{equation}
\lambda_2 =\frac{\sqrt{18 A \xi _0^5+3 F_{10} \xi _0^4+6 F_9 \xi _0^3-6 F_8 \xi _0^2-24 F_7 \xi _0+8 F_6+9 \xi _0^6}}{4 \sqrt{2} C}+\frac{3 A \xi _0^2}{4 \sqrt{2}C}+3 \beta +\frac{3 \beta  \xi _0}{2 \sqrt{2} C}+\frac{3 \xi _0^3}{4 \sqrt{2} C}+\frac{3 \xi _0}{2 \sqrt{2} C}-\frac{7}{2}
\end{equation}
\end{widetext}

where
\begin{equation}
A=\sqrt{4 \beta +\xi _0^2}
\end{equation}
\begin{equation}
C=\sqrt{A \xi _0+2 \beta +\xi _0^2}
\end{equation}
\begin{equation}
F_6=(7-6 \beta )^2 C^2-24 \beta  \left(3 \beta ^2-7 \beta +4\right)
\end{equation}
\begin{equation}
F_7=A \left(15 \beta ^2-32 \beta +16\right)+\sqrt{2} \left(-3 \beta ^2+3 \beta -1\right) C
\end{equation}
\begin{equation}
F_8=3 \sqrt{2} A (\beta -2) C+42 \beta ^2-124 \beta +58
\end{equation}
\begin{equation}
F_9=2 A (6 \beta -1)-3 \sqrt{2} (\beta -2) C
\end{equation}
\begin{equation}
F_{10}=3 A^2+24 \beta -4
\end{equation}

%
 \bibliographystyle{unsrt}
 \bibliography{viscosity}
%
%
%

\end{document}